\documentstyle[12pt]{article}
\makeatletter
\newcommand{\singlespacing}{\let\CS=\@currsize\renewcommand{\baselinestretch}{1.0}\tiny\CS}
\newcommand{\doublespacing}{\let\CS=\@currsize\renewcommand{\baselinestretch}{1.5}\tiny\CS}

\begin{document}

\title{A new $PT$ symmetric complex Hamiltonian with a real
spectra}
\author{B. Bagchi\thanks{bbagchi @ cucs.ernet.in} $^1$ \\ Department of Applied
Mathematics\\ University of Calcutta\\ 92 Acharya Prafulla Chandra Road\\
Calcutta 700009\\ India\\ \\ and\\ \\ R. Roychoudhury\thanks{raj @
www.isical.ac.in} $^2$\\ Physics \& Applied Mathematics Unit\\ Indian
Statistical Institute \\ 203 Barrackpore Trunk Road\\ Calcutta 700035\\ India}

\date{}

\maketitle

\vspace*{1cm}

\centerline{\bf Abstract}

\vspace{0.3cm}

\thispagestyle{empty}

\setlength{\baselineskip}{18.5pt}

We construct an isospectral system in terms of a real and a complex potential
to show that the underlying $PT$ symmetric complex Hamiltonian possesses
a real spectra which is shared by its real partner.

\newpage
Complex potentials have found a wide usage [1] in the literature especially
in connection with scattering problems.
Recently it has been emphasized [2] that by enforcing a $PT$-symmetry, one can
obtain new classes of complex Hamiltonian which exhibit a real spectra of
energy eigenvalues. 
  The purpose of this letter is to bring to light a new complex 
Hamiltonian which is $PT$ symmetric and possess a real energy spectra.

Consider potentials of the form $V^{(1), (2)} = U^2 \pm U^\prime$ where $U$ is
complex function of $x$ and a dash denotes a derivative with respect to $x$.
Let us express $U$ explicitly as $a(x) + ib(x)$, where $a(x)$ and $b(x)$ are
certain real, continuously differentiable functions in $R$. We have ,
$$\displaystyle{V^{(1), (2)} ~=~ (a^2 - b^2 \pm a^\prime) + i (2ab \pm
b^\prime)} \eqno {(1)}$$

In the following we investigate the possibility when one of the potentials
defined by (1) is real but the other is complex. To this end we choose,
for the sake of concreteness, $V^{(2)}$ to be real thus resticting the
function a to be given by 
$$\displaystyle{a ~=~ \frac{1}{2} ~ \frac{b^\prime}{b}} \eqno {(2)}$$
The consrtaint (2) gives for $V^{(1)}$ and $V^{(2)}$ the forms 
$$\displaystyle{V^{(1)} ~=~ (a^2 - b^2 + a^\prime) + 2 i b^\prime} \eqno
{(3a)}$$ 
$$\displaystyle{V^{(2)} ~=~ (a^2 - b^2 - a^\prime)} \eqno {(3b)}$$ 

Notice that $V^{(1)} = V^{(2)} + 2 U^\prime$ which seems to hint at a
supersymmetric connection between $V^{(1)}$ and $V^{(2)}$; however, it should
be borne in mind that the Hamiltonians for $V^{(1)}$ and $V^{(2)}$ cannot be
made simultaneously hermitian.

Our task now is to demonstrate by appropriately choosing the functions $a$ and
$b$ that the Hamiltonians $H^{(1), (2)} = - \frac{d^2}{dx^2} + V^{(1), (2)}$
yield a common real spectra thus  forming an isospectral system. We propose the
following example
$$\displaystyle{a ~=~ - \frac{\mu}{2} ~ tanh~ \mu x, ~~~ b ~=~ \lambda ~ sech~
\mu x} \eqno {(4)}$$ 
where $\mu$ and $\lambda$ are non-zero real parameters with $\mu \neq
\lambda$. Note that the above representations of the functions $a$ and $b$ are
consistent with the requirement (2).

Substitution of  (4) in (3) gives for $V^{(1)}$ and $V^{(2)}$ the expressions 
$$\displaystyle{V^{(1)} ~=~ \frac{\mu^2}{4} - \mu^2 \left[ \bar{\lambda}
(\bar{\lambda}-1) +1 \right] ~sech^2 \mu x - 2 i \lambda \mu ~sech~\mu x
~tanh~ \mu x} \eqno {(5)}$$ 
$$\displaystyle{V^{(2)} ~=~ \frac{\mu^2}{4} - \mu^2 ~\bar{\lambda}
~(\bar{\lambda}-1) ~sech^2 \mu x} \eqno {(5b)}$$ 
where $\bar{\lambda} (\bar{\lambda}-1) ~=~ \frac{\lambda^2}{\mu^2} -
\frac{1}{4}$. 

It is well known [3] that the non-zero energy levels for $V^{(2)}$
are given by
$$\displaystyle{E_n^{(2)} ~=~ \frac{\mu^2}{4} - ( \bar{\lambda} - 1 - n)^2
\mu^2, ~ n < \bar{\lambda} - 1} \eqno {(6)}$$
with $n = 0, 1, \cdots$ . The associated eigenfunctions for the even and odd
states are
$$\displaystyle{\psi_{even}^{(2)} (x) ~=~ cosh^{\bar{\lambda}} \mu x ~_{2}F_1
\left[ \frac{1}{2} (\bar{\lambda}-1), \frac{1}{2} (\bar{\lambda}+1),
\frac{1}{2} - sinh^2 \mu x \right]} \eqno {(7)}$$
$$\displaystyle{\psi_{odd}^{(2)} (x) ~=~ cosh^{\bar{\lambda}} \mu x ~ sinh~
\mu x ~_{2}F_1 \left( \frac{\bar{\lambda}}{2},~ \frac{\bar{\lambda}}{2} + 1,
\frac{3}{2}, - sinh^2 \mu x \right)} \eqno {(8)}$$

The main point of this note is to expose the fact that the complex
Hamiltonian $H^{(1)}$ ,in addition to the zero-energy level, mimicks the
discrete real spectra (7) for a class of complex eigenfunctions. We illustrate
this by considering the case $\frac{\lambda}{\mu} = - \frac{5}{2}$ for which
$\bar{\lambda} = 3$. The relevant values of $n$ then are 0 and 1, none of
which, however, correspond to the zero-energy state for either $V^{(1)}$ or
$V^{(2)}$. In Table 1 we furnish our results which have been obtained
exploiting the interwining relations $\psi^{(1)}_{(n=0,1)} = \left(\frac{d}{dx}
+ U \right) \psi^{(2)}_{(n=0,1)}$. It is obvious from there that not only  the
energy eigenvalues for $V^{(1)}$ are real and match  these of $V^{(2)}$
for $n = 0$ and 1 respectivly, but also  that the corresponding eigenfunctions for
both $V^{(1)}$ and $V^{(2)}$ have controllable  asymptotic behaviour.

Further it is clear that the complex potential $V^{(1)}$ has a normalizable 
zero-energy state with the wave-function
$$\displaystyle{\psi_0^{(1)} \alpha \sqrt{sech~\mu x}~e^{2 i \frac{\lambda}
{\mu}~tan^{-1} (e^{\mu x})}}  \eqno {(9)}$$

This is indeed a nice result considering the fact that the imaginary part of $V^{(1)}$
only contributes a phase factor in $\psi_0^{(1)}$.

To conclude, we have found a new $PT$ symmetric complex potential whose energy levels are
negative semi-definite and, excluding the zero-energy state, coinside with
those of a known $sech^2$-potential. Results for  different values of $\bar{\lambda}$ 
and other choices of $a$ and $b$ will be communicated in a future
detailed publication.

\section*{References}

\begin{enumerate}
\item[1.] See for an early work, Feshbach H,Porten C.E. and Weisskopf 
1954 Phys Rev {\bf 96}, 448.
\item[2.] Bender, C.M. and Boettcher, S., 1998 Phys. Rev. Lett. {\bf 80},
5243. 
\item[] Znojil, M., 1999 Phys. Lett. {\bf 259 A}, 220.
\item[] Bender, C.M., Cooper, F., Meisinger, P.N. and Savage, U.M., 1999
Phys. Lett {\bf 259 A}, 224.
\item[] Fernandez F.M., Guardiola Ros J and Znojil M., 1999 J. Phys. {\bf A 32},
3105.
\item[] Bender C.M. and Boettcher S., 1998 J.Phys. {\bf A 31 L} 273.
\item[] Andrianov A. A. {\it et al} Quant-ph/9806019. 
\item[3.] Fl\"{u}gge, S., 1971 Practical Quantum Mechanics, Vol {\bf 1}, New
York, Springer.
\end{enumerate}

\vspace*{0.7cm}

\noindent
TABLE 1 : \ Eigenfunctions and eigenvalues of the potentials $V^{(1), (2)}$
for the case $\bar{\lambda} = 3$.\\ 
\hspace*{2.3cm}{See text for details. $N_0$ and $N_1^{(i)}$
and normalization constants.}

\vspace{0.3cm}

\begin{tabular}{|c|c|c|c|}\hline
Eigenfunctions & For the potential $V^{(1)}$ & For the potential $V^{(2)}$ &
Energy eigenvalues\\
               & $(i = 1)$                     & $(i = 2)$     & \\ \hline
$\psi_{(n=0)}^{(i)}$ & $N_0 ~sech^2 \mu x ~(tanh~\mu x$ & $N_0 ~sech^2 \mu x$
& $\frac{\mu^2}{4} - 4 \mu^2$\\ 
               & $+ i ~ sech~\mu x)$ &             & \\ \hline
$\psi_{(n=1)}^{(i)}$ & $N_1^{(1)} ~sech \mu x ~(1 - \frac{5}{3}~sech^2\mu x$ &
$N_1^{(2)} ~sech^2 \mu x ~ sinh \mu x$ & $\frac{\mu^2}{4} - \mu^2$\\ 
               & $+ i~ \frac{5}{3} ~ sech^2 \mu x ~sinh \mu x )$ &     & \\ 
\hline
\end{tabular}

\vspace*{0.7cm}

\noindent
Footnote :\\
$1$ \ Under parity \ $(P)$ \ $p \to - p$ and $x \to - x$\\
\hspace*{1cm}{whereas under time reversal \ $(T)$ \ $p \to - p$, $x \to - x$
and $i \to - i$.
\end{document}